# Internet, Social Media and Conflict Studies
## Can Greater Interdisciplinarity Solve the Analytical Deadlocks in Cybersecurity Research?

H. Akın Ünver

Abstract: In recent years, computational research methods, digital trace data and online human interactions have contributed to the emergence of new technology-oriented sub-fields within International Relations (IR). Although the cybersecurity scholarship had an initial promise to be the *primus inter pares* among these emerging fields, the main thrust of this new methodological innovation came through the 'digital conflict studies' sub-field. By integrating Internet and social media research tools and questions into its core topics of sub-national violence, terrorism and radical mobilization, digital conflict studies has recently succeeded in addressing some of the data validity and methodology problems faced by the cybersecurity scholarship. This article begins by briefly reviewing some of the persistent data and method-oriented hurdles faced by the cybersecurity scholarship. Then, it moves onto a more detailed account of how digital conflict studies have been addressing some of these deadlocks by focusing individually on the literature on onset, mobilization, targeting, intensity/duration and termination phases of conflicts. Ultimately, the article concludes with the suggestion that the cybersecurity scholarship could move past its own deadlocks by building more granular and dedicated research datasets and establishing mechanisms to share event data with the scientific community.

## INTRODUCTION

The long-touted theoretical slumber of International Relations (IR) as a discipline[1] was markedly disturbed by the advent of the cybersecurity scholarship. Described by many scholars as the 'real fourth domain', even surpassing the importance of space, digitization of human power relations and their competition across an entirely new ground - cyber - opened up the way for the 'revival' of IR.[2] The pre-cyber IR was deadlocked for two main reasons: first, in the aftermath of the Cold War the field had lost its connection to the very origins of its existence - likelihood of large-scale globalized war. With no pressing global threat to study and theorize, the field was split across deconstructivist and post-positivist sub-strands that ran into their own set of limitations[3]. Second, the post-9/11 turn in IR had lowered the analytical lens of the bulk of the discipline into terrorism and sub-national violence, making IR more of a surrogate mother of a comparative civil war or rebel dynamics offspring. The conflict studies field has exploded so much in the last decade that it has ended up almost swallowing IR. Yet, it focused more on insurgency, micro-dynamics of violence with a heavy focus on dyadic socio-political interactions between state and non-state actors, none of which being the central thrust of the IR as a discipline.

---

[1] David A. Lake, "Theory Is Dead, Long Live Theory: The End of the Great Debates and the Rise of Eclecticism in International Relations," *European Journal of International Relations* 19, no. 3 (September 1, 2013): 567–87, https://doi.org/10.1177/1354066113494330; Rosa Vasilaki, "Provincialising IR? Deadlocks and Prospects in Post-Western IR Theory," *Millennium* 41, no. 1 (September 1, 2012): 3–22, https://doi.org/10.1177/0305829812451720; Stefano Guzzini, "The Ends of International Relations Theory: Stages of Reflexivity and Modes of Theorizing," *European Journal of International Relations* 19, no. 3 (September 1, 2013): 521–41, https://doi.org/10.1177/1354066113494327; Barry Buzan and Richard Little, "Why International Relations Has Failed as an Intellectual Project and What to Do About It," *Millennium* 30, no. 1 (January 1, 2001): 19–39, https://doi.org/10.1177/03058298010300010401.

[2] Nazli Choucri and David D. Clark, "Integrating Cyberspace and International Relations: The Co-Evolution Dilemma," SSRN Scholarly Paper (Rochester, NY: Social Science Research Network, November 7, 2012), https://papers.ssrn.com/abstract=2178586; Chintan Vaishnav, Nazli Choucri, and David Clark, "Cyber International

Cybersecurity promised salvation to IR from being sucked up into the vortex of conflict and terrorism research. It offered scholars a truly global network of state and non-state (benign and malicious) interactions pool, made some of the more classical concepts of the field (such as bandwagoning, hedging, deterrence, retaliation) relevant again and offered a new research agenda that went beyond conflict and violence. Attribution problems brought a new cyber-institutionalist tradition[4]. Attacker's advantage in cybersecurity re-introduced the question of alliances, balancing and confidence-building measures[5]. The fluid and largely uncharted nature of cyberspace reminded scholars of the importance of the anarchy and hegemony debate, as well as the centrality of the rules and norms literature in IR. While cyber hasn't yet influenced the field as much as either conventional or nuclear weapons did back in the Cold War, it has undoubtedly introduced a new wave of issue-broadening nudges, briefly widening the discipline's obsession with sub-national violence.

---

Relations as an Integrated System," *Environment Systems and Decisions* 33, no. 4 (December 1, 2013): 561–76, https://doi.org/10.1007/s10669-013-9480-3; Lucas Kello, "The Meaning of the Cyber Revolution: Perils to Theory and Statecraft," *International Security* 38, no. 2 (October 1, 2013): 7–40, https://doi.org/10.1162/ISEC_a_00138.

[3] Jonathan Havercroft and Raymond Duvall, "Challenges of an Agonistic Constructivism for International Relations," *Polity* 49, no. 1 (December 22, 2016): 156–64, https://doi.org/10.1086/689978; Jarrod Hayes, "Reclaiming Constructivism: Identity and the Practice of the Study of International Relations," *PS: Political Science & Politics* 50, no. 1 (January 2017): 89–92, https://doi.org/10.1017/S1049096516002213; David M. McCourt, "Practice Theory and Relationalism as the New Constructivism," *International Studies Quarterly* 60, no. 3 (September 1, 2016): 475–85, https://doi.org/10.1093/isq/sqw036; Thomas J. Biersteker, "Critical Reflections on Post-Positivism in International Relations," *International Studies Quarterly* 33, no. 3 (September 1, 1989): 263–67, https://doi.org/10.2307/2600459; Thomas Diez and Jill Steans, "A Useful Dialogue? Habermas and International Relations," *Review of International Studies* 31, no. 1 (January 2005): 127–40, https://doi.org/10.1017/S0260210505006339.

[4] Nicholas Tsagourias, "Cyber Attacks, Self-Defence and the Problem of Attribution," *Journal of Conflict and Security Law* 17, no. 2 (July 1, 2012): 229–44, https://doi.org/10.1093/jcsl/krs019; Thomas Rid and Ben Buchanan, "Attributing Cyber Attacks," *Journal of Strategic Studies* 38, no. 1–2 (January 2, 2015): 4–37, https://doi.org/10.1080/01402390.2014.977382.

[5] Joseph S. Nye, "Nuclear Lessons for Cyber Security?," *Strategic Studies Quarterly* 5, no. 4 (2011): 18–38.

However, cyber-research also ended up in a *cul-de-sac* of sorts. Due to the sheer magnitude of 'events' in cyber research, it brought a range of conceptual and measurement problems. These problems have led to some unforthcoming answers to essential questions about how to study the field: What constitutes a cyber 'attack'[6]? How do we disentangle offensive and defensive capabilities in cyber[7]? How do we acquire and catalogue event data, given the sheer magnitude and near-impossibility of reliably measuring cyber competition? How do we verify the accuracy of attribution? Given attribution problems, how do we reliably establish norms and minimize commitment problems in compliance? Who are we establishing norms for, or against? On the other end of the debate is the 'megafauna' argument that criticized the field's over-emphasis on a small number on politically critical cyber events (usually with physical implications), overlooking the importance of the large volume of 'smaller events' that don't have a measurable or visible impact on infrastructure or systems[8]. Both ends of the debate give us a different hurdle facing cyber-research: do we have too much, or too little measurable data to work with? All of these data availability and quality issues of cyber haven't prevented theorizing about it in IR but made empirical tests of those claims highly difficult.

Ultimately, it ran into two difficulties: first, an inability to grow beyond the confines of traditional IR and second, dangerously treading on the 'purely theoretical' territory. Even when thinking solely in terms of the disruptive effects of cyber, the nuclear analogy ran into a number of problems. After all, nuclear weapons had a binary nature – they are either detonated or not detonated. Their deterrent nature came from the possibility of detonation. Cybersecurity, on the other hand, is anything but binary. DDoS, MitM, spear phishing, SQL injection and XXS are different attack types that serve a different purpose. Often, the target learns about the fact that it was attacked weeks after the actual incident. More sophisticated forms of cyberattacks are also hard to notice, measure, turn into event data and are even harder to attribute to an actor. Its twists and turns, intensity and type all require a much more nuanced understanding of the engineering aspects of the field compared to nuclear weapons. It also rendered a truly multi-method approach, through combined forces of computer engineers, data scientists and IR specialists necessary, but this kind of sustained interdisciplinarity is still rare in the scholarship. Furthermore, persistent problems in generating a statistically meaningful (let alone representative) dataset make cybersecurity theories difficulty to test empirically.

---

[6] Scott W. Beidleman, "Defining and Deterring Cyber War" (ARMY WAR COLL CARLISLE BARRACKS PA, January 6, 2009), https://apps.dtic.mil/docs/citations/ADA500795.

[7] Rebecca Slayton, "What Is the Cyber Offense-Defense Balance? Conceptions, Causes, and Assessment," *International Security* 41, no. 3 (January 1, 2017): 72–109, https://doi.org/10.1162/ISEC_a_00267.

[8] Alexander D. Kent "Cyber security data sources for dynamic network research." In Dynamic Networks and Cyber-Security, 2016. pp. 37-65., Radu F. Babiceanu, and Remzi Seker. "Big Data and virtualization for manufacturing cyber-physical systems: A survey of the current status and future outlook." Computers in Industry 81 (2016): 128-137.

As cybersecurity scholarship dealt with these structural problems, Internet and social media studies began solving some of those persistent problems in harvesting trace data to help build new theories on technology and IR. The power of social media, both as a form of real-time communication and also a networking hub, has been demonstrated most visibly, first during the Arab Spring demonstrations in the MENA region and then in Occupy movements across Europe and North America[9]. Then with the onset and intensification of the Syrian, Iraqi and Ukrainian civil wars, the power of digital interconnectedness as an enabler of mobilization, resource generation, propaganda, and issue framing has been further bolstered[10]. The highly interactive and public nature of social media platforms not only increased the speed of communication between actors but also rendered this communication visible to the broader social media community, all around the world. Social media has become a new marketplace of ideas, equally disliked and needed, by politicians, diplomats, international institutions, civil society groups, and violent groups alike.

Compared to cybersecurity, where data is both uncontrollably large, elusive and fast, Internet research has rendered the study of human interactions in digital space more manageable. Although data size and speed can still be an issue, social media data is less elusive compared to cybersecurity data and depending on the platform and topic, scraping entire sets of public data can be far easier compared to cataloguing cyber event data. Also, the secrecy and covertness of cybersecurity research can be a deterrent factor for social scientists, which might explain why more of them are switching to the open source nature of social media research. Furthermore, although cyber research is significantly important for IR theory, it lies distant to the wider empirical debates on sample size, representativeness, measurement, causal effects, and behavioural variances[11]. Ultimately, cybersecurity research risks the danger of remaining an elite-focused, small-N field compared to the large-scale 'social' scope of social media research, until cybersecurity researchers solve the riddle of data extraction, dataset sharing and measurement. This has led to a rapidly developing field of computational social science: the use of coding and programming tools to study social dynamics, behaviour, and choice[12]. Despite enriching other sub-fields in IR, however, Internet and social media studies have primarily benefited the conflict and political violence discipline. To that end, when exploring how Internet Communication Tools (ICTs) affect IR, the bulk of the scholarship again comes from the conflict studies camp.

---

[9] Nahed Eltantawy and Julie B. Wiest. "The Arab spring| Social media in the Egyptian revolution: reconsidering resource mobilization theory." International Journal of Communication 5 (2011): 18. Philip N. Howard, Aiden Duffy, Deen Freelon, Muzammil M. Hussain, Will Mari, and Marwa Maziad. "Opening closed regimes: what was the role of social media during the Arab Spring?." Available at SSRN 2595096 (2011). Gadi Wolfsfeld, Elad Segev, and Tamir Sheafer. "Social media and the Arab Spring: Politics comes first." The International Journal of Press/Politics 18, no. 2 (2013): 115-137.

[10] Philip N . Howard and Muzammil M. Hussain. "The role of digital media." Journal of democracy 22, no. 3 (2011): 35-48. Daniel Byman and Jennifer Williams. "Jihadism's Global Civil War." The National Interest 136 (2015): 10-18. For a broader range of cases, see: Joshua Tucker, Andrew Guess, Pablo Barberá, Cristian Vaccari, Alexandra Siegel, Sergey Sanovich, Denis Stukal, and Brendan Nyhan. *Social media, political polarization, and political disinformation: A review of the scientific literature*. Hewlett Foundation report (2018).

The remainder of the paper will discuss how 'digital conflict studies' field has incorporated the 'technological turn' in IR into its main theoretical thrust, by building upon and expanding the framework used by Lars-Erik Cederman and Manuel Vogt[13] on the main phases of conflicts: onset/initiation, resource mobilization, target prioritization, duration/intensity and conflict termination, peacebuilding, re-integration. Although a similar phasing or periodization is hard to model exactly on the cybersecurity scholarship, this review aims to provide a more structured framework for the empirical study of cybersecurity. Furthermore, this periodization may help future cybersecurity scholars to structure their data observation, extraction, and sharing mechanisms to test existing IR-related cybersecurity theories and build new ones.

INTERNET, SOCIAL MEDIA AND CONFLICT: PRIMARY ANALYTICAL FOCI

As outlined in the critical review articles by Anita Gohdes[14] and Thomas Zeitzoff[15], there is quite a rapid development of robust, empirically sophisticated studies on how digital interconnectivity impacts contentious politics and human conflict. The earlier studies inspired by the Arab Spring and Occupy movements focused on the effect of digital media on protests, focusing mainly on onset and state responses against digital mobilization[16]. As protests evolved into deadlier conflicts, the field has also evolved into explaining how ICTs affected violence in civil wars like in Syria, Iraq, and Ukraine[17]. Important studies, some of which will be explored in this paper, have primarily focused on conflict initiation, recruitment, propaganda, terrorist attacks, narrative-building, branding/marketing of armed groups and state responses.

---

[11] Jean Camp, Lorrie Cranor, Nick Feamster, Joan Feigenbaum, Stephanie Forrest, Dave Kotz, Wenke Lee, Patrick Lincoln, Vern Paxson, Mike Reiter, Ron Rivest, William Sanders, Stefan Savage, Sean Smith, Eugene Spafford, and Sal Stolfo. "Data for cybersecurity research: Process and "wish list"." in: National Science Foundation Workshop on Cyber Security Data for Experimentation (2010).

[12] David Lazer et al., "Life in the Network: The Coming Age of Computational Social Science," *Science (New York, N.Y.)* 323, no. 5915 (February 6, 2009): 721–23, https://doi.org/10.1126/science.1167742; R. Conte et al., "Manifesto of Computational Social Science," *The European Physical Journal Special Topics* 214, no. 1 (November 1, 2012): 325–46, https://doi.org/10.1140/epjst/e2012-01697-8.

[13] Lars-Erik Cederman and Manuel Vogt, "Dynamics and Logics of Civil War". *Journal of Conflict Resolution*, 61, no. 9 (October, 2017): 1992-2016

[14] Anita R. Gohdes, "Studying the Internet and Violent Conflict," *Conflict Management and Peace Science*, October 25, 2017, 0738894217733878, https://doi.org/10.1177/0738894217733878.

[15] Thomas Zeitzoff, "How Social Media Is Changing Conflict," *Journal of Conflict Resolution* 61, no. 9 (October, 2017): 1970–91, https://doi.org/10.1177/0022002717721392.

However, there are several fundamental conflict dynamics that are still understudied, such as conflict duration, bargaining/negotiation and conflict termination, group disbanding and political transition of violent conflicts. Furthermore, although there are ample studies on how social media radicalizes individuals[18], there is an insufficient body of empirical investigation on how digital interconnectedness de-radicalizes or moderates users. This section is divided into five sub-sections that deal with some of the main works in their respective sub-fields and the current analytical limitations of that sub-field. These are namely: conflict onset, resource mobilization, target selection, intensity/duration, and conflict termination, peacebuilding, reintegration.

    a.    Onset: Grievance/Greed Framing

---

[16] Mohammad-Munir Adi, *The Usage of Social Media in the Arab Spring* (LIT Verlag Münster, 2014); Maximillian Hänska Ahy, "Networked Communication and the Arab Spring: Linking Broadcast and Social Media," *New Media & Society* 18, no. 1 (January 1, 2016): 99–116, https://doi.org/10.1177/1461444814538634; Nahed Eltantawy and Julie B. Wiest, "The Arab Spring| Social Media in the Egyptian Revolution: Reconsidering Resource Mobilization Theory," *International Journal of Communication* 5, no. 0 (September 2, 2011): 18; Philip N. Howard et al., "Opening Closed Regimes: What Was the Role of Social Media During the Arab Spring?," SSRN Scholarly Paper (Rochester, NY: Social Science Research Network, 2011), https://papers.ssrn.com/abstract=2595096; Habibul Haque Khondker, "Role of the New Media in the Arab Spring," *Globalizations* 8, no. 5 (October 1, 2011): 675–79, https://doi.org/10.1080/14747731.2011.621287; Axel Bruns, Tim Highfield, and Jean Burgess, "The Arab Spring and Social Media Audiences: English and Arabic Twitter Users and Their Networks," *American Behavioral Scientist* 57, no. 7 (July 1, 2013): 871–98, https://doi.org/10.1177/0002764213479374.

[17] Jytte Klausen, "Tweeting the Jihad: Social media networks of Western foreign fighters in Syria and Iraq." Studies in Conflict & Terrorism 38, no. 1 (2015): 1-22. Thomas Elkjer Nissen, "Terror. com: IS's social media warfare in Syria and Iraq." Contemporary Conflicts 2, no. 2 (2014): 2-8. James P. Farwell, "The media strategy of ISIS." Survival 56, no. 6 (2014): 49-55. Jad Melki, and May Jabado. "Mediated public diplomacy of the Islamic State in Iraq and Syria: The synergistic use of terrorism, social media and branding." Media and Communication 4, no. 2 (2016): 92-103. Olga Onuch, "EuroMaidan protests in Ukraine: Social media versus social networks." Problems of Post-Communism 62, no. 4 (2015): 217-235. , Jacob N. Shapiro and Nils B. Weidmann. "Is the phone mightier than the sword? Cellphones and insurgent violence in Iraq." *International Organization* 69 (2): 247- 274. (2015)

[18] Robin L. Thompson, "Radicalization and the Use of Social Media," *Journal of Strategic Security* 4, no. 4 (Winter 2011): 167–90; Scott Gates and Sukanya Podder, "Social Media, Recruitment, Allegiance and the Islamic State," *Perspectives on Terrorism* 9, no. 4 (2015): 107–16; Lieven Pauwels and Nele Schils, "Differential Online Exposure to Extremist Content and Political Violence: Testing the Relative Strength of Social Learning and Competing Perspectives," *Terrorism and Political Violence* 28, no. 1 (January 1, 2016): 1–29, https://doi.org/10.1080/09546553.2013.876414; Matthew Costello et al., "Who Views Online Extremism? Individual Attributes Leading to Exposure," *Computers in Human Behavior* 63 (October 1, 2016): 311–20, https://doi.org/10.1016/j.chb.2016.05.033.

Social media has expedited the dissemination of frames related to sustained political grievances and opportunities[19]. Its scope and speed led to the emergence of non-geographical communities of grievance, all receiving information about repression, disenfranchisement, and oppression from their areas of concern[20]. Similarly, and a strategy best employed by ISIS, opportunity structures in conflict, such as wealth, prestige and social acceptance are now transmitted across the world, leading to the mobilization of an unprecedented level of diversity in its recruits[21]. Of course, social media is not the only medium through which such frames have been disseminated in the past. Newspapers, novels and motion picture have all helped communicate grievance and greed narratives in the past. What makes social media novel, however, is its speed and size. Compared to older forms of media (perhaps, with the slight exception of the live broadcasting technologies), social media and the Internet offer real-time dissemination of narratives to a far larger audience. Some of the time-lag and media access (urban/rural, rich/poor) constraints that were the hallmarks of older examples of grievance/greed narrative diffusion are less relevant in the case of social media. Owing to the lowered barriers of entry, digital communication thus increases the speed and size of both the responses to framing stimuli and also the materialization of counter-narratives.

In addition to speed and scale, ICTs also offer a permanent, non-physical and omniaccessible archival function for its users[22]. This is especially valid for grievance diffusion[23], which is a crucial driver of early mobilization. By both spreading and permanently recording events and narratives in digital space, users bestow an archival power to digital media. This, in turn, allows grievances to be communicated not just quicker, but also consumed in a sustained and at-will manner. In addition to enabling masses to mobilize quickly, this sustained grievance effect of ICTs also intensifies existing social, ethnic and religious tensions, through daily access and consumption of grievance-related content[24]. This does not only impact the likelihood and frequency of conflicts but also determines the size and social support for these conflicts. In a way, the real-time archival utility of social media both drives violence onset likelihood, as well as how intense these acts of violence become. This is true both for the digital content disseminated by a single source to multiple individual users (propaganda and information cascades) and also content sought and acquired by the users themselves (information-seeking).

---

[19] Paul Collier and Anke Hoeffler, "Greed and Grievance in Civil War," Policy Research Working Paper (Washington DC: The World Bank, May 2006); Paul Collier, Anke Hoeffler, and Dominic Rohner, "Beyond Greed and Grievance: Feasibility and Civil War," *Oxford Economic Papers* 61, no. 1 (January 1, 2009): 1–27, https://doi.org/10.1093/oep/gpn029; Marc L. Hutchison, Salvatore Schiano, and Jenifer Whitten-Woodring, "When the Fourth Estate Becomes a Fifth Column: The Effect of Media Freedom and Social Intolerance on Civil Conflict," *The International Journal of Press/Politics* 21, no. 2 (April 1, 2016): 165–87, https://doi.org/10.1177/1940161216632362.

[20] Sebastián Valenzuela, Arturo Arriagada, and Andrés Scherman, "The Social Media Basis of Youth Protest Behavior: The Case of Chile," *Journal of Communication* 62, no. 2 (2012): 299–314, https://doi.org/10.1111/j.1460-2466.2012.01635.x; Sarah Anne Rennick, "Personal Grievance Sharing, Frame Alignment, and Hybrid Organisational Structures: The Role of Social Media in North Africa's 2011 Uprisings," *Journal of Contemporary African Studies* 31, no. 2 (April 1, 2013): 156–74, https://doi.org/10.1080/02589001.2013.781322; Sebastián Valenzuela, "Unpacking the Use of Social Media for Protest Behavior: The Roles of Information, Opinion Expression, and Activism," *American Behavioral Scientist* 57, no. 7 (July 1, 2013): 920–42, https://doi.org/10.1177/0002764213479375.

[21] James P. Farwell, "The Media Strategy of ISIS," *Survival* 56, no. 6 (November 2, 2014): 49–55, https://doi.org/10.1080/00396338.2014.985436; J.M. Berger, "The Metronome of Apocalyptic Time: Social Media as Carrier Wave for Millenarian Contagion," *Perspectives on Terrorism* 9, no. 4 (2015): 61–71; Anita Peresin and Alberto Cervone, "The Western Muhajirat of ISIS," *Studies in Conflict & Terrorism* 38, no. 7 (July 3, 2015): 495–509, https://doi.org/10.1080/1057610X.2015.1025611.

[22] Andrew T. Little, "Communication Technology and Protest," *The Journal of Politics* 78, no. 1 (December 17, 2015): 152–66, https://doi.org/10.1086/683187; Zeynep Tufekci and Christopher Wilson, "Social Media and the Decision to Participate in Political Protest: Observations From Tahrir Square," *Journal of Communication* 62, no. 2 (April 1, 2012): 363–79, https://doi.org/10.1111/j.1460-2466.2012.01629.x.

ICTs and conflict onset are a relatively well-studied dyad. Currently, some of the frontiers in this sub-field remain the use of fake news, trolls and bots - deliberate information manipulators - and identify how they impact greed/grievance dynamic differently than the flow of accurate information[25]. Does disinformation, for example, enable the spread of emotional and mobilizing content better than deliberate and accurate information?[26] Do non-state actors use disinformation as part of their more extensive strategic logic, or do they find disinformation as a potentially trust-mitigating factor?

Much of these research questions were tested overwhelmingly on ISIS, rendering the group as a litmus test of some of the most prominent political communication theories in digital space[27]. Yet, the scholarship that lies at the intersection of political communication and computer science fields do try to address a broader range of cases beyond ISIS. Purohit (et al.)[28] and Karuna (et al.)[29] use text-mining tools to look at how digital frames on gender-based violence are disseminated online, using a corpus-based latent Dirichlet analysis (LDA). Both studies find a greater collaboration between grievance campaigns in geographically distant communities and discover a significant degree of 'digital cross-pollination' between physically disconnected online campaigns. De De Choudhury (et al.)[30] on the other hand explore how images and videos on social media contribute to the intensification or de-sensitization of emotions in Mexico, related to cartel-based violence. They focus on sub-district level digital content across two years and find that both the state and cartels strategically deploy images to deter, and assert psychological dominance against each other as well as the local populace in a methodical way. Similarly, Monroy-Hernandez (et al.)[31] dissect to what extent 'civic bloggers' (locals who document cartel violence on weblogs and social media) have any effect on conflict awareness and mobilization. They discover that districts that contain more public media bloggers and thus, are exposed to greater volume and frequency of conflict reporting have a greater tendency to be desensitized and less likely to participate in violence.

---

[23] Qiongyou Pu and Stephen J. Scanlan, "Communicating Injustice?," *Information, Communication & Society* 15, no. 4 (May 1, 2012): 572–90, https://doi.org/10.1080/1369118X.2012.665937; Rebecca Kay LeFebvre and Crystal Armstrong, "Grievance-Based Social Movement Mobilization in the #Ferguson Twitter Storm," *New Media & Society* 20, no. 1 (January 1, 2018): 8–28, https://doi.org/10.1177/1461444816644697.

[24] Anita Breuer, Todd Landman, and Dorothea Farquhar, "Social Media and Protest Mobilization: Evidence from the Tunisian Revolution," *Democratization* 22, no. 4 (June 7, 2015): 764–92, https://doi.org/10.1080/13510347.2014.885505; Gary Tang, "Mobilization by Images: TV Screen and Mediated Instant Grievances in the Umbrella Movement," *Chinese Journal of Communication* 8, no. 4 (October 2, 2015): 338–55, https://doi.org/10.1080/17544750.2015.1086398.

[25] Meital Balmas, "When Fake News Becomes Real: Combined Exposure to Multiple News Sources and Political Attitudes of Inefficacy, Alienation, and Cynicism," *Communication Research* 41, no. 3 (April 1, 2014): 430–54, https://doi.org/10.1177/0093650212453600; Samuel C. Woolley and Philip N. Howard, "Automation, Algorithms, and Politics| Political Communication, Computational Propaganda, and Autonomous Agents — Introduction," *International Journal of Communication* 10, no. 0 (October 12, 2016): 9.

[26] For a good overview of the emerging literature on US disinformation, see: Brendan Nyhan "How Misinformation and Polarization Affect American Democracy" in Joshua Tucker, Andrew Guess, Pablo Barberá, Cristian Vaccari, Alexandra Siegel, Sergey Sanovich, Denis Stukal, and Brendan Nyhan. *Social media, political polarization, and political disinformation: A review of the scientific literature.* Hewlett Foundation report (2018).

Wang (et al.)[32] on the other hand explore the digital impact of the Sandy Hook shootings on users' sentiments on gun control, discovering that anti-violence (pro-gun-control) sentiments linger on much longer compared to anti-gun control content in the aftermath of the shooting. They also demonstrate that these pro-gun-control sentiments linger on longer in Connecticut (the incident state), compared to other states in the United States, invalidating the main premise of the 'terror management theory'. In the same vein, Ayers (et al.)[33] conduct large-scale scraping of websites, forums, social media sites and online advertisements in the US, revealing that online awareness-building campaigns and tailored ads on digital platforms could measurably mitigate gun-related deaths. While most of the 'images and frames' literature rely on text-based methods, Won (et al.)[34] focused on images shared on social media websites to draw a causal link between emotions that those images trigger and the actual intensity of the relevant protest event. Drawing on more than 40,000 geotagged protest images, the researchers find that greater protest duration and intensity are correlated with the emotional triggering capacity of the images shared from the protest area. Finally, Muller and Schwartz demonstrate a causal relationship between anti-refugee sentiment in German-speaking Facebook and physical anti-refugee violence in Germany at the municipality level[35], while Patton (et al.)[36] draw direct links 'Internet bragging' and tendency to exhibit gang violence-related behaviour. Building a predictive model, the authors demonstrate a high level of reliability in enabling social workers recognize gang violence-related themes and engage in better targeted preventive action.

---

[27] Farwell, "The Media Strategy of ISIS"; Elizabeth Pearson, "Online as the New Frontline: Affect, Gender, and ISIS-Take-Down on Social Media," *Studies in Conflict & Terrorism* 41, no. 11 (November 2, 2018): 850–74, https://doi.org/10.1080/1057610X.2017.1352280; Yair Galily et al., "The Boston Game and the ISIS Match: Terrorism, Media, and Sport," *American Behavioral Scientist* 60, no. 9 (August 1, 2016): 1057–67, https://doi.org/10.1177/0002764216632844.

[28] Hemant Purohit et al., "Gender-Based Violence in 140 Characters or Fewer: A #BigData Case Study of Twitter," *ArXiv:1503.02086 [Cs]*, March 6, 2015, http://arxiv.org/abs/1503.02086.

[29] Prakruthi Karuna et al., "On the Dynamics of Local to Global Campaigns for Curbing Gender-Based Violence," *ArXiv:1608.01648 [Cs]*, August 4, 2016, http://arxiv.org/abs/1608.01648.

[30] Munmun De Choudhury, Andrés Monroy-Hernández, and Gloria Mark, "'Narco' Emotions: Affect and Desensitization in Social Media during the Mexican Drug War," *Proceedings of the 32nd Annual ACM Conference on Human Factors in Computing Systems - CHI '14*, 2014, 3563–72, https://doi.org/10.1145/2556288.2557197.

[31] Andrés Monroy-Hernández et al., "The New War Correspondents: The Rise of Civic Media Curation in Urban Warfare," July 5, 2015, https://doi.org/10.1145/2441776.2441938.

[32] Nan Wang, Blesson Varghese, and Peter D. Donnelly, "A Machine Learning Analysis of Twitter Sentiment to the Sandy Hook Shootings," *ArXiv:1609.00536 [Cs]*, September 2, 2016, http://arxiv.org/abs/1609.00536.

These studies push the field strongly into the domain of automated-recognition and prediction foci. Most of these studies are also heavily concentrated on the US or the UK, requiring a significant broadening of the cases beyond the global north. Some of the untapped research questions include how regional inequality, linguistic or ethnic fractionalization, and regime type influence the ICT-onset literature.

### b. Resource Mobilization

Not all successful cases of greed/grievance communication lead to violence. Without pre-existing social forces and trust relations, purely digital networks have a low level of success of mobilizing or remaining mobilized for an extended period[37]. One of the main drivers of the staying power of movements is how well they mobilize resources[38]. This mobilization is not just limited to manpower or financial resources but also includes sustenance of frames, attention, alliances, commitment and organizational cohesion. To that end, the advent of social media and the Internet are primarily thought as positive drivers of resource mobilization. After all, this was one of the main reasons why the early works on the Arab Spring and Occupy movements were quick to call them 'social media revolutions'[39]. When dissidents discovered their latent power through digital interconnectivity, they were able to come together and challenge hegemony, or so the logic went. However, later studies have proved that this was not always the case[40]. Social media was a significant development but was also exaggerated in its effect on social movements. ICTs did drive attention and awareness, but these two variables had a mixed effect on actual mobilization, commitment or staying power. Rather, ICTs effects relied on pre-existing traditional networks and trust relations; it is only when these existing networks are in place that ICTs could play a major role.

---

[33] John W. Ayers et al., "Can Big Media Data Revolutionarize Gun Violence Prevention?," *ArXiv:1611.01148 [Cs]*, November 3, 2016, http://arxiv.org/abs/1611.01148.

[34] Donghyeon Won, Zachary C. Steinert-Threlkeld, and Jungseock Joo, "Protest Activity Detection and Perceived Violence Estimation from Social Media Images," *ArXiv:1709.06204 [Cs]*, September 18, 2017, http://arxiv.org/abs/1709.06204.

[35] Karsten Müller and Carlo Schwarz. "Fanning the flames of hate: Social media and hate crime." Available at SSRN 3082972 (2018).

[36] Desmond U. Patton, Jeffrey Lane, Patrick Leonard, Jamie Macbeth, and Jocelyn R. Smith Lee. "Gang violence on the digital street: Case study of a South Side Chicago gang member's Twitter communication." *New Media & Society* 19, no. 7 (2017): 1000-1018.

[37] Zeynep Tufekci, Twitter and tear gas: The power and fragility of networked protest. Yale University Press, 2017. José Van Dijck, Thomas Poell, and Martijn De Waal. The platform society: Public values in a connective world. Oxford University Press, 2018. Bernard Enjolras, Kari Steen-Johnsen, and Dag Wollebæk. "Social media and mobilization to offline demonstrations: Transcending participatory divides?." New media & Society 15, no. 6 (2013):

This necessitates a greater empirical focus on the analytical sweet spot that delves into how traditional social networks use and deploy ICTs to generate sustained resource. Mostly, studies omit this context and try to draw a straightforward causal link between how much movements use social media and how well they mobilize. Furthermore, past the onset and mobilization mark, few studies explore how movements sustain both ideological and material resources, especially when the rival actor is also a non-state actor[41]. To that end, most scholarship focuses on how protestors, rioters, militias and terrorist groups utilize ICTs against state actors, but not much on how this interaction materializes between and within non-state actors themselves. Some of the most novel ways of studying mobilization dynamics can be seen in Gahot (et al.)[42] where the authors use epidemiological modelling to see the diffusion and sustenance of the 2005

---

890-908. Anita Breuer, Todd Landman, and Dorothea Farquhar. "Social media and protest mobilization: Evidence from the Tunisian revolution." Democratization 22, no. 4 (2015): 764-792.

[38] Kristen Lovejoy and Gregory D. Saxton, "Information, Community, and Action: How Nonprofit Organizations Use Social Media," *Journal of Computer-Mediated Communication* 17, no. 3 (April 1, 2012): 337–53, https://doi.org/10.1111/j.1083-6101.2012.01576.x; Breuer, Landman, and Farquhar, "Social Media and Protest Mobilization"; Enjolras, Steen-Johnsen, and Wollebæk, "Social Media and Mobilization to Offline Demonstrations."

[39] Khondker, "Role of the New Media in the Arab Spring"; Bruns, Highfield, and Burgess, "The Arab Spring and Social Media Audiences"; Christopher Wilson and Alexandra Dunn, "The Arab Spring| Digital Media in the Egyptian Revolution: Descriptive Analysis from the Tahrir Data Set," *International Journal of Communication* 5, no. 0 (September 2, 2011): 25.

[40] Eltantawy and Wiest, "The Arab Spring| Social Media in the Egyptian Revolution"; Michael Hoffman and Amaney Jamal, "Religion in the Arab Spring: Between Two Competing Narratives," *The Journal of Politics* 76, no. 3 (July 1, 2014): 593–606, https://doi.org/10.1017/S0022381614000152.

[41] Anita R. Gohdes, "Studying the Internet and Violent Conflict."

[42] Laurent Bonnasse-Gahot et al., "Epidemiological Modelling of the 2005 French Riots: A Spreading Wave and the Role of Contagion," *Scientific Reports* 8, no. 1 (January 8, 2018): 107, https://doi.org/10.1038/s41598-017-18093-4.

French riots. This is one of the better experiments in bringing greater interdisciplinarity into social research, as the authors pay specific attention to social dynamics (neighbourhood relations, district social networks, and protestor community proximity) in their contagion model. In a similar vein, Davies (et al.)[43] explore how 2011 London riots have been mobilized against police crackdown, using digital communication tools to build a contagion dataset of both onset and sustained resource-generation dynamics. Ultimately, the authors discover that there is a Pareto-optimal level of police presence per each riot district that deters rioter resource mobilization; any less or more police deployment causes greater protest mobilization and resources deployed in the riot.

ICT-religion nexus is a potentially helpful disciplinary link to conflict studies, especially in terms of exploring how loose belief networks retain membership, income and resource-generation practices in digital space. Heidi Campbell demonstrates how Christian congregations that are geographically distant can retain 'online congregations' through live-feed sermons and chat groups[44]. These online congregations not only build awareness and disseminate frames, but also play a significant role in the material domain, in terms of fundraising and mustering numbers for protests and social responsibility projects[45]. In the same vein, Ellison and Boyd illustrate how this 'online congregation' dynamic can be observed in Islam and Judaism as well, as individuals discover and often times identify with their digital congregations more than their physical ones[46]. Mellor and Rinnawi focus specifically on the Islamic use of ICTs and exemplify how digital congregations are more important drivers of resource mobilization in diaspora Islamic communities in Europe and the United States, whereas in predominantly Muslim countries too, the effect of social media leads to the emergence of diverging loyalties[47]. Such diverging loyalties create an '*à la carte* Islam' where the pious can choose from different sermons and fatwas issued by the online or offline imams, demonstrating hybrid loyalties based on which congregation or imam supply the version of theological interpretation they require. On Judaism, Nathan Abrams demonstrates how social media leads to 'post-denominational Judaism', where identity construction and mobilization dynamics display similar congregational hybridity that we see in Islam[48].

---

[43] Toby P. Davies et al., "A Mathematical Model of the London Riots and Their Policing," *Scientific Reports* 3 (February 21, 2013): 1303, https://doi.org/10.1038/srep01303.

[44] Heidi A. Campbell, *Digital Religion: Understanding Religious Practice in New Media Worlds* (New York, NY: Routledge, 2012), 58; Heidi A. Campbell, "Understanding the Relationship between Religion Online and Offline in a Networked Society," *Journal of the American Academy of Religion* 80, no. 1 (March 1, 2012): 64–93, https://doi.org/10.1093/jaarel/lfr074.

[45] Heidi Campbell, *When Religion Meets New Media* (Routledge, 2010), 126, https://doi.org/10.4324/9780203695371.

[46] Nicole B. Ellison and Danah M. Boyd, "Sociality Through Social Network Sites," in *The Oxford Handbook of Internet Studies*, ed. William H. Dutton, Oxford Handbooks (Oxford, UK: Oxford University Press, 2013), http://www.oxfordhandbooks.com/view/10.1093/oxfordhb/9780199589074.001.0001/oxfordhb-9780199589074-e-8.

[47] Noha Mellor and Khalil Rinnawi, *Political Islam and Global Media: The Boundaries of Religious Identity* (Routledge, 2016), 126, 135.

### c. Target Prioritization

Once movements attain critical mass and are directed in unison towards a common goal, the next step becomes target selection. In the case of non-violent movements, this target becomes a key building, central square, or a historic site to occupy or do a sit-in, whereas in violent movements these range from military installations/checkpoints, critical infrastructure or (with terrorist groups) civilian areas. What is the role of social media on how groups pick their targets and guide their followers to that target? Some of the earlier works on this question had a slight reporting bias around areas of high Internet connectivity[49], although newer forms of scholarship have begun developing unique ways to disentangle violence occurrence from violence reporting[50]. Some of the most prominent studies on violence targeting emphasize four trends: strategic area/zone focus, critical infrastructure focus, political focus and intimidation focus. There is currently no major work on how ICTs affect any particular focus. Rather, most target-related studies explore how social media sustains and intensifies ongoing conflict dynamics, evidenced best in Thomas Zeitzoff's work on the 2012 Gaza Conflict[51]. Also, Jibon Naher and Matiur Rahman Minar look at how social media posts drive chat groups or page followers to particular types of violence, including the target(s) of such acts in Bangladesh[52]. They demonstrate how ICTs not only impact the speed and size of violent mobilization, but also drive users towards definitive ethnic or religious groups and sites that are important for their adversaries.

---

[48] Nathan Abrams, "Appropriation and Innovation: Facebook, Grassroots Jews and Offline Post-Denominational Judaism," in *Digital Judaism: Jewish Negotiations with Digital Media and Culture*, ed. Heidi A. Campbell (New York, NY: Routledge, 2015), https://doi.org/10.4324/9781315818597-7.

[49] Jan H. Pierskalla and Florian M. Hollenbach, "Technology and Collective Action: The Effect of Cell Phone Coverage on Political Violence in Africa," *American Political Science Review* 107, no. 2 (May 2013): 207–24, https://doi.org/10.1017/S0003055413000075; Allan Dafoe and Jason Lyall, "From Cell Phones to Conflict? Reflections on the Emerging ICT–Political Conflict Research Agenda," *Journal of Peace Research* 52, no. 3 (May 1, 2015): 401–13, https://doi.org/10.1177/0022343314563653.

[50] Nils B. Weidmann, "A Closer Look at Reporting Bias in Conflict Event Data," *American Journal of Political Science* 60, no. 1 (2016): 206–18, https://doi.org/10.1111/ajps.12196; Mihai Croicu and Joakim Kreutz, "Communication Technology and Reports on Political Violence: Cross-National Evidence Using African Events Data," *Political Research Quarterly* 70, no. 1 (March 1, 2017): 19–31, https://doi.org/10.1177/1065912916670272.

[51] Thomas Zeitzoff, "Does Social Media Influence Conflict? Evidence from the 2012 Gaza Conflict," *Journal of Conflict Resolution* 62, no. 1 (January 1, 2018): 29–63, https://doi.org/10.1177/0022002716650925.

[52] Jibon Naher and Matiur Rahman Minar, "Impact of Social Media Posts in Real Life Violence: A Case Study in Bangladesh," *ArXiv:1812.08660 [Cs]*, December 19, 2018, http://arxiv.org/abs/1812.08660.

Targeting presents a very specific causal problem in ICT-conflict nexus. The majority of existing studies can explain how ICT use can guide mobilization, intensity, polarization and radicalization, but there is not much on whether the popularity of particular target types on ICTs really affect physical targeting behaviour on the ground. This is a challenging puzzle to solve, because it is hard to disaggregate targeting choice as a result of its social media popularity, or the order of a commanding superior[53]. Further problems arise in determining whether ICTs influence commanders, or troops on the ground in hand picking a particular set of targets. Especially with regard to communal violence, the role of ICTs in instigating certain ethnic or religious groups using targeted messaging and imagery is under-explored. Similar to military targeting, there is much to be done on exploring whether the decision to attack a particular target or community follows a strict hierarchy (i.e. communal elders or propagandists) or generate a mob mentality (autonomous crowd behaviour). Some tangential journalism work on this is done through WhatsApp-related gang violence; specifically, in India[54]. Some of the most relevant findings on this line of inquiry are that digital targeting practices through text or images contribute substantially to the unearthing of dormant social tensions, leading to attacks on women specifically. This gender dimension in violent group targeting prioritization is an important avenue for further research and connects somewhat to another strand of literature on digital shaming and harassment[55]. Furthermore, due to the real-time effects of ICTs, there is far less time gap between designating a target in digital space and the actual mobilization of the masses around the designated target.

I have discovered this targeting effect of ICTs when exploring the mobilization dynamics during the failed coup in Turkey in July 2016[56]. Sharing videos, images, and text in large numbers, anti-coup protestors were extremely quick in mobilizing and overwhelming checkpoints, military installations and other hotspots of conflict. In most cases, it is possible to draw direct linkages between the emergence of a particular hotspot or target on social media and the massing of crowds around it within an hour or less. We also, however, warn against the existing social network dimension of these targeting preferences. It is not only ICTs that guided people to specific locations; it is rather the dissemination of these targets through family, friend or religious congregation channels (all trusted pre-existing networks) through ICTs that created the trust network required to march towards designated spots. Future ICT-targeting studies have to take this digital-physical interaction into account.

---

[53] Lisa Hultman, Jacob Kathman, and Megan Shannon, "United Nations Peacekeeping and Civilian Protection in Civil War," *American Journal of Political Science* 57, no. 4 (2013): 875–91, https://doi.org/10.1111/ajps.12036; Jessica A. Stanton, "Regulating Militias: Governments, Militias, and Civilian Targeting in Civil War," *Journal of Conflict Resolution* 59, no. 5 (August 1, 2015): 899–923, https://doi.org/10.1177/0022002715576751.

[54] Vindu Goel, Suhasini Raj, and Priyadarshini Ravichandran, "How WhatsApp Leads Mobs to Murder in India," *The New York Times*, July 18, 2018, sec. Technology, https://www.nytimes.com/interactive/2018/07/18/technology/whatsapp-india-killings.html, https://www.nytimes.com/interactive/2018/07/18/technology/whatsapp-india-killings.html; Marcos Martínez, "Burned to Death Because of a Rumour on WhatsApp," *BBC News*, November 12, 2018, sec. Latin America & Caribbean, https://www.bbc.com/news/world-latin-america-46145986.

[55] Jessica Megarry, "Online Incivility or Sexual Harassment? Conceptualising Women's Experiences in the Digital Age," *Women's Studies International Forum* 47 (November 1, 2014): 46–55, https://doi.org/10.1016/j.wsif.2014.07.012; Jesse Fox and Wai Yen Tang, "Women's Experiences with General and Sexual Harassment in Online Video Games: Rumination, Organizational Responsiveness, Withdrawal, and Coping Strategies," *New Media & Society* 19, no. 8 (August 1, 2017): 1290–1307, https://doi.org/10.1177/1461444816635778.

[56] H. Akin Unver and Hassan Alassaad, "How Turks Mobilized Against the Coup," *Foreign Affairs*, September 14, 2016, https://www.foreignaffairs.com/articles/2016-09-14/how-turks-mobilized-against-coup.

### d. Intensity and Duration

Conflict intensity and duration have grown increasingly popular in the last decade, owing to new datasets and measurement types allowing greater analytical focus. The primary thrust of the conflict intensity/duration literature has been to inform policy on the ways of shortening conflicts.[57] By reducing conflict duration, the conflict literature asserted, conflict intensity would also be reduced[58]. However, premature freezing of conflicts due to external pressure also frequently lead to a re-escalation of those conflicts once international attention wanes[59]. Therefore, the causal relationship between duration and intensity is still being negotiated in the literature as this link works differently in various political contexts. Although conflict intensity (measured as casualties) and duration (measured as months) have been explained most popularly through combatant motivation, financing, public commitment, capacity and external support[60], the role of ICTs has been a relatively nascent inquiry. The common consensus in the conflict literature is that ICTs make conflicts more intense[61]. This finding is being increasingly challenged by a new scholarship that focuses on the availability bias and measurement level problems of these findings. Nils Weidmann, for example, asserts convincingly that access to ICTs creates a reporting bias[62]. This reporting bias is both related to the communication of violent cases that would remain uncommunicated without ICTs and also that those with access to ICTs tend to over-share violent events compared to non-violent events in a conflict zone. This brings the need for interdisciplinarity to solve some of the availability bias and data-gathering problems related to proximity to ICT infrastructure.

---

[57] E. Elbadawi and N. Sambanis, "Why Are There so Many Civil Wars in Africa? Understanding and Preventing Violent Conflict," *Journal of African Economies* 9, no. 3 (October 1, 2000): 244–69, https://doi.org/10.1093/jae/9.3.244; Ibahim Elbadawi and Nicholas Sambanis, "How Much War Will We See?: Explaining the Prevalence of Civil War," *Journal of Conflict Resolution* 46, no. 3 (June 1, 2002): 307–34, https://doi.org/10.1177/0022002702046003001.

[58] Megan Shannon, Daniel Morey, and Frederick J. Boehmke, "The Influence of International Organizations on Militarized Dispute Initiation and Duration," *International Studies Quarterly* 54, no. 4 (December 1, 2010): 1123–41, https://doi.org/10.1111/j.1468-2478.2010.00629.x.

[59] Elena Pokalova, "Conflict Resolution in Frozen Conflicts: Timing in Nagorno-Karabakh," *Journal of Balkan and Near Eastern Studies* 17, no. 1 (January 2, 2015): 68–85, https://doi.org/10.1080/19448953.2014.986378; Stephen Blank, "Russia and the Black Sea's Frozen Conflicts in Strategic Perspective," *Mediterranean Quarterly* 19, no. 3 (August 28, 2008): 23–54.

[60] Håvard Hegre, "The Duration and Termination of Civil War," *Journal of Peace Research* 41, no. 3 (May 1, 2004): 243–52, https://doi.org/10.1177/0022343304043768; Karl R. de Rouen and David Sobek, "The Dynamics of Civil War Duration and Outcome," *Journal of Peace Research* 41, no. 3 (May 1, 2004): 303–20, https://doi.org/10.1177/0022343304043771; James D. Fearon, "Why Do Some Civil Wars Last So Much Longer than Others?," *Journal of Peace Research* 41, no. 3 (May 1, 2004): 275–301, https://doi.org/10.1177/0022343304043770.

Guo (et al.)[63] for example, demonstrate how betweenness centrality can offer a higher explanatory power on why certain geographies become more conflict-prone than others. They emphasize cities as hubs of communication and thus, bestow greater importance to the cities for their narrative and frame-dissemination capacity that in turn influence conflict intensity and duration patterns. In their view, conflict intensity and duration are both proportional to their distance to major cities, and also the betweenness centrality measures of major cities themselves. Walther (et al.)[64] on the other hand focus on the Sahel-Sahara to aggregate digital and pre-digital communication and archival data to generate a militant attack database. Using this database, they are able to infer which national communication and counter-insurgency policies worked better in terms of mitigating the intensity and duration of the Al Qaeda in Mali (AQIM) violence in the Sahel region. However, with the exception of these few examples, the ICT-duration or intensity literature is still very young. A potential analytical field to approach the intensity/duration problem would be to engage more with the digital political communication literature.

Echo chambers and polarization are proper proxy measures to solve this deadlock. Barberá (et. al)[65] demonstrate how political preferences are disseminated across accounts that are similar in political ideology, but that such similarity disappears in non-political issues. This could be an inspirational point of departure for conflict studies as the dissemination of conflict-related information along ideological or identity lines would potentially have a significant influence on conflict intensity. Del Vicaro (et al.)[66] on the other hand, demonstrate how disinformation is especially contingent upon network ideology. Focusing on how conspiracy theories flow on Facebook, authors demonstrate that polarization becomes more observable in the dissemination of false narratives rather than accurate information. This is especially relevant for conflict research, as emotion-eliciting information tends to be fake and it is mostly fake news that provokes violent action among groups. Bastos (et al.)[67] test the same hypothesis in the Brexit campaign and discover that both accurate and inaccurate types of content are shared across a more dense and central network in the Leave camp and follows a looser and less dense network in the Stay camp. This finding is important as it hints at the role of political ideology in the dissemination of accurate and false-type conflict-related content. In a more recent study, Cota (et al.)[68] look at a large dataset of Twitter messages from Brazil, related to the impeachment of President Dilma Rousseff and discover that more extreme content got disseminated across a broader audience, compared to less extreme and more moderate content that had a low level of spread. This study sheds some light on potential conflict communications literature on the degree to which extreme versus moderate content becomes popular during emergencies, escalations and critical political moments.

---

[61] T Camber Warren, "Explosive Connections? Mass Media, Social Media, and the Geography of Collective Violence in African States," *Journal of Peace Research* 52, no. 3 (May 1, 2015): 297–311, https://doi.org/10.1177/0022343314558102; Dinissa Duvanova et al., "Violent Conflict and Online Segregation: An Analysis of Social Network Communication across Ukraine's Regions," *Journal of Comparative Economics*, Ukraine\: Escape from Post-Soviet Legacy, 44, no. 1 (February 1, 2016): 163–81, https://doi.org/10.1016/j.jce.2015.10.003.

[62] Weidmann, "A Closer Look at Reporting Bias in Conflict Event Data."

[63] Weisi Guo et al., "The Spatial Ecology of War and Peace," *ArXiv:1604.01693 [Physics]*, April 6, 2016, http://arxiv.org/abs/1604.01693.

[64] Olivier Walther, Christian Leuprecht, and David Skillicorn, "Wars Without Beginning or End: Violent Political Organizations and Irregular Warfare in the Sahel-Sahara," *ArXiv:1606.02705 [Physics]*, June 8, 2016, http://arxiv.org/abs/1606.02705.

[65] Pablo Barberá et al., "Tweeting From Left to Right: Is Online Political Communication More Than an Echo Chamber?," *Psychological Science* 26, no. 10 (October 1, 2015): 1531–42, https://doi.org/10.1177/0956797615594620.

### e. Termination, Peacebuilding, Re-integration

How do ICTs impact the decision to end a conflict? The literature is skewed so much on the initiative and sustaining dynamics of ICTs on conflict, that the analytical 'wave' hasn't yet reached the final stages of the violence cycle. This owes partly to the fact that recent years have witnessed much more conflict initiation and escalation compared to conflict termination[69]. Most conflict termination scholarship focuses on tangible factors, such as the decline of manpower, funds or other resources to explain de-escalation and termination dynamics[70]. Other studies explain how external interventions, infighting, leadership change and social shifts might lead to similar outcomes[71]. However, there have been no empirical studies yet on how ICTs impact the de-escalation and termination aspects of conflicts.

---

[66] Michela Del Vicario et al., "Public Discourse and News Consumption on Online Social Media: A Quantitative, Cross-Platform Analysis of the Italian Referendum," *ArXiv:1702.06016 [Physics]*, February 20, 2017, http://arxiv.org/abs/1702.06016; Michela Del Vicario et al., "Polarization and Fake News: Early Warning of Potential Misinformation Targets," *ArXiv:1802.01400 [Cs]*, February 5, 2018, http://arxiv.org/abs/1802.01400.

[67] Marco T. Bastos, Dan Mercea, and Andrea Baronchelli, "The Spatial Dimension of Online Echo Chambers," *ArXiv:1709.05233 [Physics]*, September 15, 2017, http://arxiv.org/abs/1709.05233.

[68] Wesley Cota et al., "Quantifying Echo Chamber Effects in Information Spreading over Political Communication Networks," *ArXiv:1901.03688 [Physics]*, January 11, 2019, http://arxiv.org/abs/1901.03688.

[69] Nils-Christian Bormann, Lars-Erik Cederman, and Manuel Vogt, "Language, Religion, and Ethnic Civil War," *Journal of Conflict Resolution* 61, no. 4 (April 1, 2017): 744–71, https://doi.org/10.1177/0022002715600755.

Earlier hypotheses on how ICTs might influence conflict offered promising notes. Most of them thought of greater interconnectedness and at-will information availability to have a normalizing effect on violent interactions. Importing from more traditional theories on media and conflict, the early ICT literature suggested that warring factions or mobilized masses could be 'convinced' to stand down, and attain greater mutual understanding with the rival groups (or governments), resulting in conflict termination[72]. Most of the 1990s' peacebuilding initiatives thus emphasized 'greater communication' and frequent socialization between the sides and by 2000s, most such initiatives welcomed the idea of digital communication as a peace enabler[73]. The evidence yielded a more complex and nuanced picture, however. One main misjudgment about ICT-peace link was the fact that ICTs grew into multi-directional media systems, as opposed to the uni-directional nature of TV, motion picture, or radio. This meant that 'feeding content' to warring factions, in the same manner propaganda worked in the past, was impossible[74]. Content consumers are now also content producers and curators, causing far too many narratives to spread online for any single actor to dominate the mainstream agenda at-will[75]. The prevalence of multiple and competing narratives, along with the ever-present shadow of disinformation also have a chance of intensifying existing divisions, rather than healing them[76]. To counter these effects, along with attempts to skew these discussions by combing out 'unwanted content', governments have also begun increasingly reliant on censorship, net restrictions, as well as the strategic deployment of bots and trolls to further contaminate the information ecosystem[77]. However, these tools are still not dominantly in use in most conflict-prone countries as instability prevents the establishment of functioning Internet infrastructure. This severs the link between ICTs and affected communities in rural areas or poor urban districts, and renders any uni-directional move towards mediation inefficient.

---

[70] Joakim Kreutz, "How and When Armed Conflicts End: Introducing the UCDP Conflict Termination Dataset," *Journal of Peace Research* 47, no. 2 (March 1, 2010): 243–50, https://doi.org/10.1177/0022343309353108; J. Michael Quinn, T. David Mason, and Mehmet Gurses, "Sustaining the Peace: Determinants of Civil War Recurrence," *International Interactions* 33, no. 2 (April 20, 2007): 167–93, https://doi.org/10.1080/03050620701277673; David E. Cunningham, Kristian Skrede Gleditsch, and Idean Salehyan, "It Takes Two: A Dyadic Analysis of Civil War Duration and Outcome," *Journal of Conflict Resolution* 53, no. 4 (August 1, 2009): 570–97, https://doi.org/10.1177/0022002709336458.

[71] Monica Duffy Toft, "Ending Civil Wars: A Case for Rebel Victory?," *International Security* 34, no. 4 (March 17, 2010): 7–36, https://doi.org/10.1162/isec.2010.34.4.7; Barbara F. Walter, "Does Conflict Beget Conflict? Explaining Recurring Civil War," *Journal of Peace Research* 41, no. 3 (May 1, 2004): 371–88, https://doi.org/10.1177/0022343304043775; Michael Tiernay, "Killing Kony: Leadership Change and Civil War Termination," *Journal of Conflict Resolution* 59, no. 2 (March 1, 2015): 175–206, https://doi.org/10.1177/0022002713499720.

[72] Devon E. A. Curtis, "Broadcasting Peace: An Analysis of Local Media Post-Conflict Peacebuilding Projects in Rwanda and Bosnia," *Canadian Journal of Development Studies / Revue Canadienne d'études Du Développement* 21, no. 1 (January 1, 2000): 141–66, https://doi.org/10.1080/02255189.2000.9669886; Wilhelm Kempf, "Media Contribution to Peace Building in War Torn Societies," Working Paper, Diskussionsbeiträge Der Projektgruppe Friedensforschung (Konstanz: University of Konstanz, 1998), https://kops.uni-konstanz.de/handle/123456789/10701.

[73] Monroe E. Price and Mark Thompson, "Intervention, Media and Human Rights," *Survival* 45, no. 1 (March 1, 2003): 183–202, https://doi.org/10.1093/survival/45.1.183; Gadi Wolfsfeld, Eitan Y. Alimi, and Wasfi Kailani, "News Media and Peace Building in Asymmetrical Conflicts: The Flow of News between Jordan and Israel," *Political Studies* 56, no. 2 (June 1, 2008): 374–98, https://doi.org/10.1111/j.1467-9248.2007.00683.x.

[74] Clay Shirky, "The Political Power of Social Media: Technology, the Public Sphere, and Political Change," *Foreign Affairs* 90, no. 1 (2011): 28–41.

From existing empirical evidence, it is hard to hypothesize that ICTs will have any standalone effect on conflict termination and peacebuilding, without interacting with material factors on the ground. Some of these material factors are already well-covered by the non-ICT conflict termination literature: external intervention, combatant resolve, leadership change, resource depletion, social change and so on. In an interaction with any of these factors, ICTs may bolster peacekeeping efforts, provided that the material factors are realistically in place and in a visible, sustained manner.

The final link in the conflict chain is attaining sustained non-violence and community-building in the affected territories. Currently, there are no significant works in the field of how ICTs impact peacekeeping and community-building, leaving a sizeable analytical gap in the literature. This originates from the overwhelming focus on the earlier onset/mobilization aspects of violence, leaving a plethora of untapped topics on the later phases of violence. Yet, the literature will inevitably gravitate towards the more positive and constructive aspects of ICTs on conflict. Community-building and re-integration phase of the conflict cycle have a higher likelihood of being impacted by purely ICT-led efforts. When willingness to fight is diminished and the societal-political consensus is in place, then ICT tools can work through a multitude of ways to help bolster the peace agreement and its enforcement. Most direct and visible use of ICTs in this phase is to monitor a ceasefire and other commitment violations, enabling all parties to keep an eye on the other in a more direct way[78]. When the speed and scale advantages of ICTs are deployed after peace is attained, then these assets can significantly bolster confidence-building measures and render community-building and re-integration efforts more robust. In other words, ICTs will work, when there are 'physical' attempts towards reconciliation, and will remain ineffective without any social consensus or driver towards mediation.

---

[75] Oliver Boyd-Barrett, "Ukraine, Mainstream Media and Conflict Propaganda," *Journalism Studies* 18, no. 8 (August 3, 2017): 1016–34, https://doi.org/10.1080/1461670X.2015.1099461.

[76] Nadiya Kostyuk and Yuri M. Zhukov, "Invisible Digital Front: Can Cyber Attacks Shape Battlefield Events?," *Journal of Conflict Resolution* 63, no. 2 (February 1, 2019): 317–47, https://doi.org/10.1177/0022002717737138; Theodore P. Gerber and Jane Zavisca, "Does Russian Propaganda Work?," *The Washington Quarterly* 39, no. 2 (April 2, 2016): 79–98, https://doi.org/10.1080/0163660X.2016.1204398; Ulises A Mejias and Nikolai E Vokuev, "Disinformation and the Media: The Case of Russia and Ukraine," *Media, Culture & Society* 39, no. 7 (October 1, 2017): 1027–42, https://doi.org/10.1177/0163443716686672.

[77] Gary King, Jennifer Pan, and Margaret E. Roberts, "How Censorship in China Allows Government Criticism but Silences Collective Expression," *American Political Science Review* 107, no. 2 (May 2013): 326–43, https://doi.org/10.1017/S0003055413000014; Yi Mou, Kevin Wu, and David Atkin, "Understanding the Use of Circumvention Tools to Bypass Online Censorship," *New Media & Society* 18, no. 5 (May 1, 2016): 837–56, https://doi.org/10.1177/1461444814548994; Anita R. Gohdes, "Pulling the Plug: Network Disruptions and Violence in Civil Conflict," *Journal of Peace Research* 52, no. 3 (May 1, 2015): 352–67, https://doi.org/10.1177/0022343314551398; Howard et al., "Opening Closed Regimes."

## CONCLUSION

This article has aimed to trace some of the current analytical fronts of digital conflict studies to offer a framework for IR's cybersecurity scholarship on how it could escape some of the short-term methodological and conceptual deadlocks. Most directly, the cybersecurity field suffers from an absence of granular and representative (or even indicative) data generation and sharing mechanism for the purposes of the scholarly community. This prevents an empirical testing of the majority of the claims made in cybersecurity theory, as well as impair how much we can import from the oft-mentioned 'nuclear analogy' in IR for the purposes of the cybersecurity scholarship. One of the main reasons why digital conflict studies discipline could surmount most of the hurdles posed by the emerging technologies is that the field has been building granular and representative event datasets since the 1960s. From the founding blocks of the Correlates of War Project in the 1960s, UCDP/PRIO and START-UMD datasets emerged, leading up the way for the emergence of many other dataset projects including ICEWS, ACLED, ViEWS and GDELT projects. Multiple long-term, open source scientific research datasets have rendered the conflict literature more advantageous in adapting to the requirements of the technological turn in IR. Absent of a long-term, scientific event dataset, cybersecurity researchers of IR have to work with an ambiguous sample of cybersecurity events that will structurally remain insufficient for robust theory-building and theory testing.

---

[78] David Patrikarakos, *War in 140 Characters: How Social Media Is Reshaping Conflict in the Twenty-First Century* (New York: Basic Books, 2017); Christian Christensen, "Twitter Revolutions? Addressing Social Media and Dissent," *The Communication Review* 14, no. 3 (July 1, 2011): 155–57, https://doi.org/10.1080/10714421.2011.597235; Matthew L. Williams et al., "Policing Cyber-Neighbourhoods: Tension Monitoring and Social Media Networks," *Policing and Society* 23, no. 4 (December 1, 2013): 461–81, https://doi.org/10.1080/10439463.2013.780225; Anita R. Gohdes, "Studying the Internet and Violent Conflict."

A second major difference is that the digital conflict studies has grown far more interdisciplinary and collaborative than the cybersecurity scholarship. By focusing on human trace data to infer human conflict dynamics, the digital conflict literature has largely retained the social and human prerogatives of the field. In contrast, cybersecurity scholarship is still very much distant to the core questions of political and social fundamentals of digitalized human interactions, so far refraining from directly dealing with the role of psychology, sociology and ethnographic aspects of cybersecurity. One recent exception is Nadiya Kostyuk and Yuri Zhukov's exploration of how cyber-attacks influence conflict dynamics on the front lines in Ukraine.[79] This article provides one of many potential trajectories for the cybersecurity scholarship in addressing the data availability and validity issues.

ICTs bring a set of welcome challenges and opportunities for conflict research. However, the most immediate task for the IR discipline in general is to escape yet another conflict studies entrapment when successfully adapting to the question of how emerging technologies impact world events. ICTs bring a wide array of analytical opportunities to test IR's existing mainstream theories, as well as develop new ones as digital communication enables the harvesting of both more abundant and more granular data types. These data types will not only improve our understanding of how states interact with their societies and other states in the digital domain, but will also lead the way to the formulation of more advanced types of international institutions and cooperation architectures. Regardless, conflict studies will likely continue to spearhead the IR's broader disciplinary focus on the ICTs.

While there is quite an abundance of attention on the earlier phases of social movements and violent events, there is currently a gap in the more advanced phases of such phenomena. The field is still very much nascent in explaining how ICT use affects target selection, intensity/duration, and peacebuilding efforts, but have a relatively more robust set of findings on the onset and mobilization phases. To remedy this gap, digital political communication literature seems to be the most apparent linkage for the conflict studies community. Some of the most prominent works in *PolCom* already have theoretical answers to how violence affects attention, emotion and mediation in critical political events and debates.

Yet, the bulk of that literature focuses on American or European contexts, where both Internet penetration and social media use are significant enough to yield robust

---

[79] Nadiya Kostyuk and Yuri M. Zhukov, "Invisible Digital Front: Can Cyber Attacks Shape Battlefield Events?" *Journal of Conflict Resolution.* 63, no. 2 (March 2019): 317-347

results[80]. The challenge of the conflict studies literature is double-pronged: first, tackle data availability and parity issues in comparative works on countries with uneven access to ICTs, and second, disentangle conflict reporting frequency from conflict occurrence frequency. Both problems may eventually solve themselves as global interconnectivity widens and ICT data becomes more representative and comparable. However, studies can still benefit significantly from exploring how yearly, generational or geographical changes in ICT use impact all five phases of conflicts discussed above. Rather than waiting for ICT data to become eventually representative, scholars may find a plethora of topics to dig into across cases in areas with different levels of digital access. Greater interdisciplinarity with the field of computer science, physics, biology and complexity fields will enable conflict researchers to generate more sophisticated models and more robust findings on this emerging frontier.

Another largely untapped topic is the role of technology companies in driving/mitigating conflict. Platform architecture, as well as increased relevance of big tech companies during crises and emergencies, render them de facto actors in all conflicts[81]. Direct effects of platform influence on conflicts, such as the decision to censor and emphasize certain types of content, or decision to partner with the governments or certain civil society actors will have an increasing influence on all phases of conflicts. However, indirect effects, such as algorithmic bias[82], attention economy[83] and rent-generation structures of social networks[84] will also impact conflicts and conflict actors significantly. Given the fact that platforms run on advertisement revenue and those advertisements target engagement metrics, it will be inevitable for platforms to push

---

[80] A further line of inquiry would be whether heavy scholarly and policy attention on disinformation in Western liberal democracies render disinformation and organized distraction easier and more likely in countries outside NATO and/or are not liberal democracies. For a range of time-frequency analyses on countries beyond mainstream scholarly focus, see: Rolf Fredheim-led 'Robotrolling' reports available at: https://www.stratcomcoe.org/robotrolling-20171

[81] Anne Helmond, "The Platformization of the Web: Making Web Data Platform Ready," *Social Media + Society* 1, no. 2 (July 1, 2015): 2056305115603080, https://doi.org/10.1177/2056305115603080; Jean-Christophe Plantin et al., "Infrastructure Studies Meet Platform Studies in the Age of Google and Facebook," *New Media & Society* 20, no. 1 (January 1, 2018): 293–310, https://doi.org/10.1177/1461444816661553.

[82] Safiya Umoja Noble, *Algorithms of Oppression: How Search Engines Reinforce Racism*, 1 edition (New York: NYU Press, 2018).

[83] Zeynep Tufekci, "'Not This One': Social Movements, the Attention Economy, and Microcelebrity Networked Activism," *American Behavioral Scientist* 57, no. 7 (July 1, 2013): 848–70, https://doi.org/10.1177/0002764213479369; Zeynep Tufekci, "Engineering the Public: Big Data, Surveillance and Computational Politics," *First Monday* 19, no. 7 (July 2, 2014), https://doi.org/10.5210/fm.v19i7.4901; Bernard J. Jansen et al., "Twitter Power: Tweets as Electronic Word of Mouth," *Journal of the American Society for Information Science and Technology* 60, no. 11 (2009): 2169–88, https://doi.org/10.1002/asi.21149.

[84] David Lazer, "The Rise of the Social Algorithm," *Science* 348, no. 6239 (June 5, 2015): 1090–91, https://doi.org/10.1126/science.aab1422; Gunn Sara Enli and Eli Skogerbø, "Personalized Campaigns in Party-Centred Politics," *Information, Communication & Society* 16, no. 5 (June 1, 2013): 757–74, https://doi.org/10.1080/1369118X.2013.782330.

more extreme content that elicits an emotional response and thus, generate more revenue.

The success of the emerging 'digital conflict studies' can offer three main lessons for the burgeoning cybersecurity and international relations field. First, 'cyberIR' has to prioritize building event datasets that strike a balance between data size and quality. Given the fact that conflict studies has several important datasets and data processing projects unique to the field (UCDP/PRIO[85], ICEWS[86], ACLED[87], UMD-START[88], or ViEWS[89]), the field is better-suited to incorporate digital media and communication-related data types. Although Center for Strategic and International Studies[90] and Council on Foreign Relations[91] have their IR-specific 'major events' data, the cases are a very unhelpfully small fraction of cybersecurity events that happen on a daily basis. Furthermore, both datasets bring in the question of 'major event according to whom?' given the heavy bias in favour of attacks against American infrastructure. Even data from Norse Corporation, FireEye, Kaspersky, Fortinet or similar cybersecurity companies are by no means representative, or transparent in terms of the methodology behind data generation. Furthermore, their full data aren't available for public use, which brings its own set of replication problems. Second, Cyber-IR scholarship has to reach beyond the confines of traditional IR theories and begin engaging more with the techno-sociology and techno-politics literature. How do states and societies perceive and prioritize the dangers of threats originating from the cybersecurity/defense domain? What is the impact of organizational, political and social differences between and within countries in politicizing and securitizing cyber threats? How do we incorporate disinformation, information overload and digital distraction into the IR's threat spectrum? How does the 'modes of production' of digital hardware, software and data impact balance of power between countries and alter state-society relations in a global scale? Should the field deal primarily with the threats faced by states (infrastructure damage, national secrecy and confidential data hacking) or the society (surveillance, data capitalism, private data protection)? Finally, the cybersecurity scholarship will have to broaden its focus of interest beyond what the major powers are doing and study the impact of major power rivalry on the rest of the world, with specific emphasis on inter-state and sub-state inequalities. Given the fact that access to advanced cybersecurity technology is a distinct privilege of only the wealthiest of nations, it is expected to create another form of hegemonic balancing and counter-balancing behaviour. How do we study regional and systemic inequalities in access to cybersecurity hardware and software infrastructure and how do these inequalities shape state behaviour and state-society relationship across different regimes and security cultures?

---

[85] UCDP/PRIO Armed Conflict Dataset: https://www.prio.org/Data/Armed-Conflict/UCDP-PRIO/
[86] 'The Integrated Conflict Early Warning System (ICEWS)'. Lockeed Martin: https://www.lockheedmartin.com/en-us/capabilities/research-labs/advanced-technology-labs/icews.html
[87] The Armed Conflict Location & Event Data Project: https://www.acleddata.com/
[88] National Consortium for the Study of Terrorism and Responses to Terrorism: https://www.start.umd.edu/
[89] 'Violence Early-Warning System (ViEWS)'. Uppsala University: https://www.pcr.uu.se/research/views/
[90] 'Significant Cyber Incidents'. Center for Strategic and International Studies: https://www.csis.org/programs/cybersecurity-and-governance/technology-policy-program/other-projects-cybersecurity
[91] 'Cyber Operations Tracker'. Council on Foreign Relations: https://www.cfr.org/interactive/cyber-operations

Ultimately, ICTs will occupy a more important place in the media-conflict nexus compared to the historical effects of the telegram, radio, TV or motion picture. This is because access to ICTs is more intimate (i.e. through smartphones that are always nearby), real-time, multi-directional and at-will compared to all previous forms of mass communication. This will also render ICT a key analytical topic in all aspects of human relations, including conflict. More focus on the communicative aspects of mobilization, movements and violence will ultimately unlock new ideas and theories of human behavior. To do this, however, this article argued in favor of the analytical value of greater interdisciplinary and more risk-taking across fields.